\documentclass{jfm}

\usepackage{graphicx}
\usepackage{natbib}

\ifCUPmtlplainloaded \else
  \checkfont{eurm10}
  \iffontfound
    \IfFileExists{upmath.sty}
      {\typeout{^^JFound AMS Euler Roman fonts on the system,
                   using the 'upmath' package.^^J}%
       \usepackage{upmath}}
      {\typeout{^^JFound AMS Euler Roman fonts on the system, but you
                   dont seem to have the}%
       \typeout{'upmath' package installed. JFM.cls can take advantage
                 of these fonts,^^Jif you use 'upmath' package.^^J}%
      }
  \else
  \fi
\fi

\ifCUPmtlplainloaded \else
  \checkfont{msam10}
  \iffontfound
    \IfFileExists{amssymb.sty}
      {\typeout{^^JFound AMS Symbol fonts on the system, using the
                'amssymb' package.^^J}%
       \usepackage{amssymb}%

      }{}
  \fi
\fi

\ifCUPmtlplainloaded \else
  \IfFileExists{amsbsy.sty}
    {\typeout{^^JFound the 'amsbsy' package on the system, using it.^^J}%
     \usepackage{amsbsy}}
    {}
\fi

\newsavebox{\astrutbox}
\sbox{\astrutbox}{\rule[-5pt]{0pt}{20pt}}

\usepackage{bm}
\usepackage{subfigure}
\usepackage{psfrag}
\usepackage{color}

\title[Where do small, weakly inertial particles go in a turbulent flow?]{Where do small, weakly inertial particles go in a turbulent flow?}

\author[M. Gibert, H. Xu and E. Bodenschatz]%
{M\ls A\ls T\ls H\ls I\ls E\ls U\ns G\ls I\ls B\ls E\ls R\ls T$^{1,4}$%
\thanks{Current address Institut N\'EEL CNRS/UJF (Grenoble France)}
\thanks{Email address for correspondence: mathieu.gibert@grenoble.cnrs.fr},\ns
H\ls A\ls I\ls T\ls A\ls O\ns X\ls U$^{1,4}$\thanks{Email address for correspondence: Haitao.Xu@ds.mpg.de},\ns\break
\and E\ls B\ls E\ls R\ls H\ls A\ls R\ls D\ns B\ls O\ls D\ls E\ls N\ls S\ls C\ls H\ls A\ls T\ls Z$^{1,2,3,4}$\thanks{Email address for correspondence: Eberhard.Bodenschatz@ds.mpg.de},\ns}

\affiliation{$^1$Max Planck Institute for Dynamics and Self Organization (MPIDS), 37077 G\"ottingen, Germany\\[\affilskip]
$^2$Institute for Nonlinear Dynamics, University of G\"ottingen, 37077 G\"ottingen, Germany\\[\affilskip]
$^3$Laboratory of Atomic and Solid-State Physics and Sibley School of Mechanical and Aerospace Engineering, Cornell University, Ithaca, New York 14853, USA\\[\affilskip]
$^4$International Collaboration for Turbulence Research}

\pubyear{2011}
\volume{650}
\pagerange{119--126}
\date{?; revised ?; accepted ?. - To be entered by editorial office}

\begin{document}

\maketitle

\begin{abstract}
We report experimental results on the dynamics of heavy particles of the size of the Kolmogorov-scale in a fully developed turbulent flow. 
The mixed Eulerian structure function of two-particle velocity and acceleration difference vectors $\langle\delta_{r}\mathbf{v}\cdot\delta_{r}\mathbf{a_{p}}\rangle$ was observed to increase significantly with particle inertia for identical flow conditions. We show that this increase is related to a preferential alignment between these dynamical quantities. With increasing particle density the probability for those two vectors to be collinear was observed to grow.  We show that these results are consistent with the preferential sampling of strain-dominated regions by inertial particles. 
\end{abstract}

\section{Introduction}
Turbulence occurs whenever fluid viscous forces are small compared to the dominant driving forces of the flow. In practice this includes most macroscopic natural and technological flows. Very frequently, these turbulent flows are loaded with small, passive particles with non-negligible inertia, \textit{i.e.} particles do not exactly follow fluid motion. Familiar examples of such flows include clouds in the atmosphere, sand storms in arid regions, flow driven transports of grains in pharmaceutical and food industries, fluidized beds in chemical engineering, and fuel sprays in combustion engines and turbines. 
The understanding of particle-turbulence interaction is therefore of central importance to our ability to make advances in key economical and societal issues like energy generation, climate change, and pollution control. 
Comparing to the fluid turbulence problem itself, which is well described by the non-linear Navier-Stokes equations, the particle-turbulence interaction problem is on much weaker foundations. 
In principle, the particle dynamics can be obtained by solving particle equations coupled with the Navier-Stokes equations for the fluid with no-slip boundary conditions on particle surfaces. However, this approach is a formidable task for theoretical and numerical investigations~\citep{happel:1965,Homann:submitted}.
Most theoretical analyses and numerical simulations therefore must rely on simplified equations, in which the particle-turbulence interactions are modeled.  An interesting result obtained from these simplified equations is that in a turbulent flow particles with small to moderate inertia and sizes smaller than the Kolomogrov length scale tend to move to regions with high strain and low vorticity~\citep{MaxeyRiley}. This so-called ``preferential concentration'' of inertial particles is under intensive investigation~\citep{Falkovich:2004p987,Zaichik:2009p3754,Chun:2005p3763,Collins:2004p3860,Sundaram:1997p3616,Ducasse:2008p3197} and has been invoked when studying droplet collision rates in warm clouds~\citep{Shaw:2003p769,falkovich:2002} and plankton encounter rates in the ocean~\citep{schmitt:2007}. Inhomogeneous distributions of inertial particles in turbulent flows have also been observed in several experiments~\citep{Aliseda:2002p3883,Wood:2005p3884,Salazar:2008p962,Saw:2008p719,Monchaux:2010}.
Here we show direct experimental evidence that regions of high particle concentrations are also regions of high strain.


\section{Experimental method \& observation}
\begin{figure}
\begin{center}
  \includegraphics[width=7.4cm]{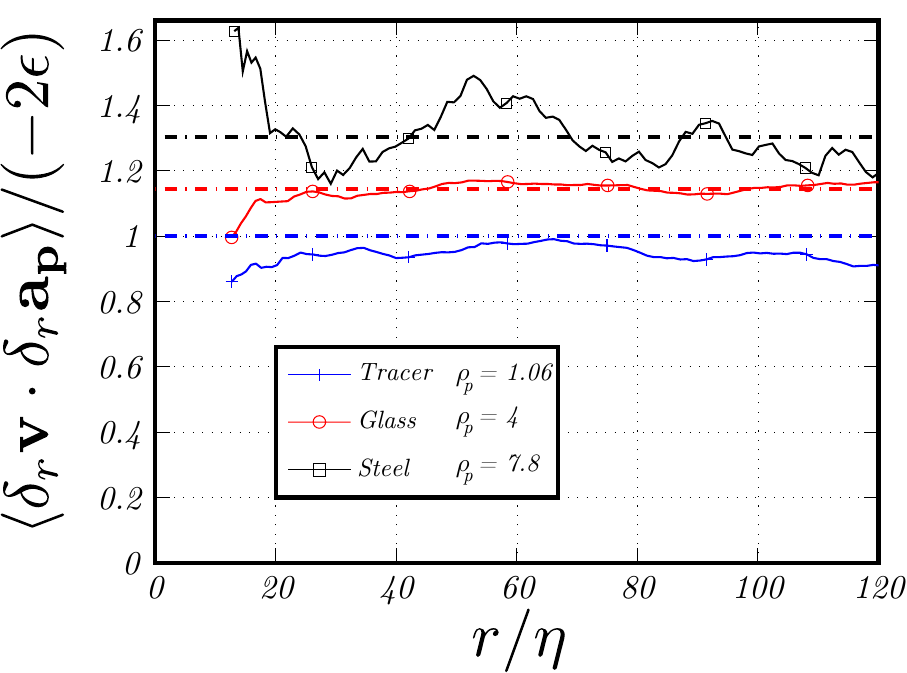}
\end{center}
\caption{\label{deltaVdeltaA}Mixed Eulerian velocity-acceleration structure function $\langle\delta_{r}\mathbf{v}\cdot\delta_{r} \mathbf{a}\rangle$ normalized by the expected values for the fluid particles $-2\epsilon$. The values of the different plateaus represented by the dot-dashed lines are : blue $1$, red $1.15$, black $1.30$. }
\end{figure}

We conducted three-dimensional Lagrangian Particle Tracking measurements~\citep{Ouellette:2006p384,Xu:2008p1921} and obtained the trajectories of solid particles with different densities in a turbulent water flow. Below we give a brief summary of the experiment (see also ~\cite{EPL_particle_2010}). 
The turbulent flow was generated by two counter-rotating baffled discs of $28\, cm$ in diameter. The turbulence chamber, shaped as a hexagonal cylinder, measured $40\, cm$ along the axis of the propellers and $38\, cm$ in the cross-section, both in height (vertically) and width (horizontally).
The axis of rotation of the discs was in the horizontal direction. The strong sweeping of the fluid near the bottom surface resuspended the heavy particles with the effect that the particle concentration remained sufficient for the measurement of multi-particle statistics.  A spherically shaped measurement volume with a diameter of $22\,mm$ was located at the center of the water chamber, where the mean flow was small compared to the velocity fluctuations. The spatial resolution of $45\, \mu$m per pixel and the temporal resolution of $31$ frames per Kolmogorov time $\tau_\eta$  guaranteed accurate measurements of particle velocity and acceleration \citep{Xu:2007p370}.
As shown in the Table \ref{tab:ExperimentParam} we used three different types of particles: polystyrene, glass, and stainless steel with densities between $1\lessapprox\rho_{p}/\rho_{f}\lessapprox8$. For the Taylor microscale Reynolds number of the turbulent flow ($R_{\lambda}=442$), the particles had the size of the  Kolmogorov scale ($\eta =74\, \mu$m).  The resulting Stokes numbers, a relative measure of the particle inertia,  were between $0.08$ and $0.5$ (see Table \ref{tab:ExperimentParam}). 
Other relevant parameters of the turbulent flow are the fluctuating velocity $u'=0.15\, m/s$, the integral length-scale $L=87\, mm$ and the Kolmogorov time-scale $\tau_{\eta}=5.2\, ms$.
We measured $3\times 10^7$ data points for the polystyrene particles, $8\times 10^7$ for the glass particles and $3\times 10^6$ for the steel particles.The fluctuations in Fig.~\ref{deltaVdeltaA} can be attributed to the statistical uncertainty.

\begin{table}
\begin{center}
\begin{tabular}{c | c | c | c | c}
 & $\rho_{p}/\rho_{f}$ & $d_p$ ($\mu$m) & $d_{p}/\eta$ & $St$\\
\hline
\hline
Polystyrene & $1.06$ & $74 \pm 10$ & $0.98$ & $0.08$\\ 
\hline
Glass & $4$ & $75 \pm 8$ & $1.04$ & $0.24$\\ 
\hline
Steel & $7.8$ & $75 \pm 15$ & $1.04$ & $0.44$\\ 
\end{tabular}
\end{center}
\caption{Characteristics of the different particles used in the experiments. The Stokes number is defined as $St \equiv d_{p}^{2}/12\beta\eta^{2}$, where $d_{p}$ is the particle diameter, $\beta \equiv 3\rho_{f}/(2\rho_{p}+\rho_{f})$ is the modified density ratio, that takes into account the added mass effect, and $\eta$ is the Kolmogorov length scale of the turbulence.}
\label{tab:ExperimentParam}
\end{table}

For these heavy particles, we observed that the magnitude of the mixed velocity-acceleration Eulerian structure function $\langle\delta_{r}\mathbf{v}\cdot\delta_{r}\mathbf{a_{p}}\rangle$ increased with particle density, where $\delta_r \mathbf{v}$ and $\delta_r \mathbf{a_p}$ are the relative velocity and the relative acceleration between two particles separated at distance $r$
(\textit{e.g.} $\delta_r \mathbf{v} = \mathbf{v}(x+r,t)-\mathbf{v}(x,t)$).
For fluid tracers this mixed velocity-acceleration structure function in the inertial range equals a constant: $\langle\delta_{r}\mathbf{u}\cdot\delta_{r}\mathbf{a}\rangle=-2\epsilon$, where $\mathbf{u}$ and $\mathbf{a}$ are fluid velocity and acceleration, respectively, and $\epsilon$ is energy dissipation rate per unit mass ~\citep{Mann:1999p3829,Falkovich:2001p4080,Pumir:2001p2611,Hill:2006}. We also found this in our experiment, where the polystyrene particles behaved as tracers \citep{EPL_particle_2010}.
As shown in Figure~\ref{deltaVdeltaA}, for inertial particles, this mixed velocity-acceleration structure functions in the inertial range were still constant, but the values increased significantly (up to $30$\%) when the particle Stokes number increased from $0.08$ to $0.44$. 

As reported earlier~\citep{EPL_particle_2010} the relative velocities in this range changed  by less than 10\%. In addition, the RMS acceleration changed with particle density from $6.9$, $6.6$ and $8.5$ m/s$^2$ for tracer, glass and steel particles, respectively.
Therefore, the observed change in $\langle\delta_{r}\mathbf{v}\cdot\delta_{r}\mathbf{a_{p}}\rangle$ cannot be simply explained by changes in the magnitudes of $\delta_r \mathbf{v}$ and $\delta_r \mathbf{a_p}$, but must be associated with the alignment between the two vectors.  Indeed, we observed a change of the cosines 
\begin{equation}
\cos \theta = \frac{\delta_{r}\mathbf{v}\cdot\delta_{r}\mathbf{a_{p}}} {|\delta_{r}\mathbf{v}||\delta_{r}\mathbf{a_{p}}|}.
\label{DefCosTheta}
\end{equation}
of the angle between these two vectors with inertia of the particles.  
Considering the tracer particles (polystyrene), its average value $\langle \cos\theta \rangle$ monotonically increases from $-0.052$ to $-0.01$ when increasing the observation scale $r$ ($12<r/\eta<120$). For glass and steel particles, the average $\langle \cos\theta \rangle$ increased in absolute value by 25\% and 9\%, respectively.

\section{Alignment of relative acceleration and relative velocity}
\begin{figure}
\begin{center}
\includegraphics[width=14cm]{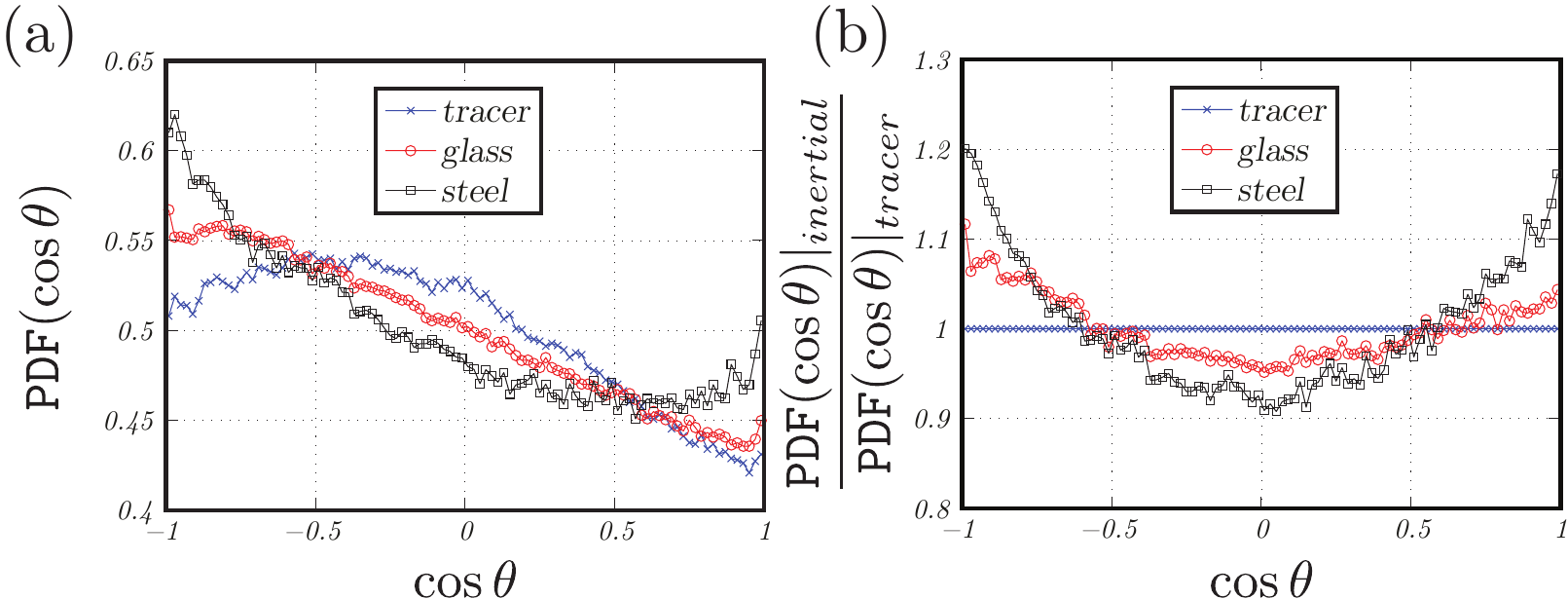}
\end{center}
\caption{\label{cosPDFexp}(a) Probability density function of the cosine of the angle $\theta$ between $\delta_{r}\mathbf{v}$ and $\delta_{r}\mathbf{a_{p}}$ for the three types of particles for a separation $r/\eta=28 \pm 7$. (b) Same PDFs as in (a) normalized by the PDF of the tracer particles. (b) shows the relative effect of the particle inertia on the alignment/anti-alignment of $\delta_{r}\mathbf{v}$ with $\delta_{r}\mathbf{a_{p}}$.}
\end{figure}

The effect of particle inertia on $\cos\theta$ is best demonstrated by its probability density function (PDF). 
Figure~\ref{cosPDFexp}a shows the PDFs of $\cos\theta$ corresponding to $\delta_r \mathbf{v}$ and $\delta_r \mathbf{a_p}$ at separation $r = (28 \pm 7)\eta$ for the three different types of particles (similar behaviour is observed for other separations in the inertial range).
All PDFs are skewed to the side of $\cos\theta < 0$, indicating that the two vectors are preferentially oriented in opposite directions, which gives the negative values of $\langle\delta_{r}\mathbf{v}\cdot\delta_{r}\mathbf{a_{p}}\rangle$, as shown in Figure~\ref{deltaVdeltaA}.

A very interesting observation from Figure~\ref{cosPDFexp}a is the increase of the probability of $|\cos\theta| \approx 1$ with particle inertia. This can be seen even better in Figure~\ref{cosPDFexp}b where the PDFs of the two types of inertial particles were normalized by the PDF of the fluid tracers. Clearly, the particle inertia enhanced the collinearity, either in the same direction or in the opposite direction, between the relative velocity and the relative acceleration vectors.  We note that the PDFs are slightly skewed toward the anti-alignment side ($\cos\theta<0$).  

 How can we understand this observation?  For this range of Stokes numbers, as shown numerically and experimentally \citep{Bec:2006p656,SalazarCollins:2010,EPL_particle_2010}, the difference between particle and fluid velocity and acceleration at the particle position is very small. The same is true for the dynamical variables $\delta_r \mathbf{a_p}$ and $\delta_r \mathbf{v}$.  Thus, if the inertial particles were homogeneously distributed in space, like fluid tracers, we should not observe any differences between the PDFs of $\cos\theta$ for fluid tracers and inertial particles. This is not what we observed. Therefore our  measurements strongly indicate  that inertial particles preferentially explore certain regions of the flow, leading to the observed PDFs in Figure~\ref{cosPDFexp}.

\section{Implication to preferential sampling}
To further elucidate  which features of the turbulence are responsible for the observed behaviour, we investigate $\delta_{r}\mathbf{u}$ and $\delta_{r}\mathbf{a}$ of fluid tracers for simple flow fields. Let us consider a simple model, namely a steady, incompressible linear velocity field $u_i$ defined by : 
\begin{equation}
u_{i} = M_{ij}x_{j}
\label{LinearFlowField}
\end{equation}
where $u_{i}$ is the fluid velocity at position $x_{i}$, $M_{ij}$ is the velocity gradient tensor. In our experiments, where the separation $r$ are in the inertial range, we could treat $M_{ij}$ as an effective velocity gradient at scale $r$. 
It has been shown that the effective velocity gradient at scale $r$ evolves with a time scale $t_0 \equiv (r^2/\epsilon)^{1/3}$~\citep{XPB2011,PXB2011}, which is much larger than the particle relaxation time $\tau_{p}$ for $r$ in the inertial range, as it can be quantified by noticing that the ratio $\tau_p / t_0 = St (\eta / r)^{2/3} \ll 1$.
Therefore, to study the dynamics of particle relative velocity and relative acceleration, the flow field may be treated as frozen.

For a steady flow, the fluid acceleration field $a_i$ is determined by $M_{ij}$: $a_{i}=u_{j}\frac{\partial u_{i}}{\partial x_{j}} = M_{ik}M_{kj}x_{j}$. 
The expressions for two-particle relative velocity and acceleration, together with the cosine of the angle between them, $\cos\theta$, can then be obtained in closed form.
For a velocity field with pure rotation, $M_{ij} = - M_{ji}$, it can be easily shown that $\delta_r \mathbf{a} \cdot \delta_r \mathbf{u} = 0$, \textit{i.e.} the fluid acceleration is always normal to its velocity. The PDF of $\cos\theta$ is a $\delta$-function at $\cos\theta = 0$. 
This is not consistent with our observations.

\begin{figure}
\begin{center}
  \includegraphics[width=7.4cm]{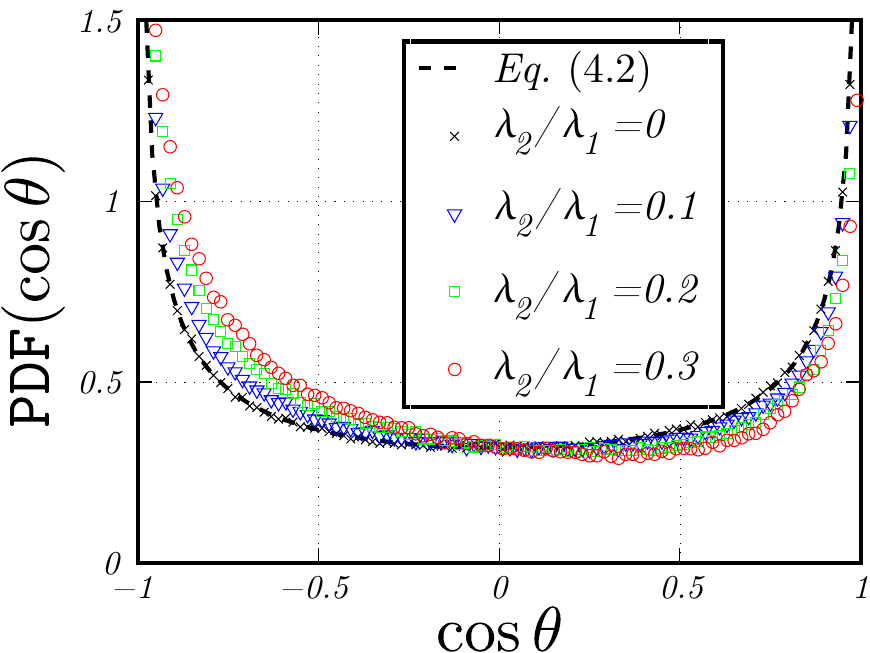}
\end{center}
\caption{\label{cosPDFth} PDF of $\cos\theta$ as defined in Eq. \ref{DefCosTheta}. The relative velocity and acceleration are obtained from two fluid tracers randomly distributed in a steady, purely straining linear velocity field. The different symbols corresponds to different ratios of $\lambda_{2}/\lambda_{1}$. The dashed-line corresponds to Eq. \ref{eq:Pcostheta}.}
\end{figure}

For a purely straining velocity field, the three eigenvalues of $M_{ij}$ are all real and can be arranged as $\lambda_{1}\geqslant\lambda_{2}\geqslant\lambda_{3}$. Incompressibility requires that $\lambda_1 + \lambda_2 + \lambda_3 = 0$. 
Then it is straightforward to show that the alignment between relative velocity and relative acceleration vectors depends only on the orientation of the separation vector $\mathbf{r}$ between the two particles and the ratio of the eigenvalues, e.g., $\lambda_2 / \lambda_1$. 
The PDF of $\cos\theta$ at a given $\lambda_{2}/\lambda_{1}$ can then be obtained by assuming that $\mathbf{r}$ distributes uniformly in all directions within this purely straining flow.
When $\lambda_2 = 0$, \textit{i.e.} the two-dimensional case, the PDF is:
\begin{equation}
\mathtt{PDF}(\cos\theta) = \frac{1}{\pi (1-\cos^{2}\theta)^{1/2}} .
\label{eq:Pcostheta}
\end{equation}
For more general cases, $\lambda_{2}/\lambda_{1}\neq 0$, we obtained the PDFs numerically by  choosing randomly $10^{7}$ separation vectors $\mathbf{r}$ and computing $\cos\theta$ (Eq.~\ref{DefCosTheta}) in the linear velocity field $u_{i}$ (Eq.~\ref{LinearFlowField}). 
Please note, numerical analysis of  various test cases showed that the shape of the PDF was insensitive to specifics of the particle distribution.
It is well known that in turbulent flows the average of the eigenvalue ratio of the velocity gradient is positive, with a value of $\langle \lambda_{2}/\lambda_{1} \rangle \approx 0.2$~\citep{BETCHOV:1956p3652,ASHURST:1987p3749,Kholmyansky:2001p3877,Luthi:2005p3878,Berg:2006p3882}.  
For the effective velocity gradient in the inertial range, this ratio decreases monotonically from this positive value to $0$ when the scale $r$ increases from $\eta$ to the integral scale~\citep{PXB2011}. 
We therefore performed simulations with the range of $\lambda_2 / \lambda_1$ around these values.
Figure~\ref{cosPDFth} shows the PDFs of $\cos\theta$ from the simulation results with $0 \leqslant  \lambda_{2}/\lambda_{1} \leqslant 0.3$, together with the analytical solution Eq.~\ref{eq:Pcostheta}. 
The PDFs for $\lambda_2 / \lambda_1 >0$ in Fig.~\ref{cosPDFth} are remarkably similar to the ``relative PDFs'' shown in Figure~\ref{cosPDFexp}b, namely, the relative accelerations and the relative velocities prefer to be collinear, with slight skewness to the anti-alignment side. This observation, together with the fact that inertial particles have, to leading order, the velocities and accelerations of the fluid at the particle positions \citep{Bec:2006p656,SalazarCollins:2010,EPL_particle_2010}, suggests that  inertial particles sample  preferentially strain-dominated regions in turbulence. 
Our approach should  be compared with results from numerical simulations of simplified equations, for which the phenomenon of ``preferential sampling of strain-dominated regions'' has been observed~\citep{Chun:2005p3763,Collins:2004p3860}. There the  velocity fields around the virtual inertial point particles was investigated.  It will be interesting to see how well the  PDFs of $\cos\theta$ agree  with the experimental results presented here.
 Moreover, studying the PDF of $\cos\theta$ for fluid tracers might reveal new information about the structures of the turbulent flow itself, \textit{e.g.} by conditioning of  the PDF on local strain rates and vorticities.

\section{Summary}
In summary,  we observed a significant increase of the mixed velocity-acceleration structure function $\langle\delta_{r}\mathbf{v}\cdot\delta_{r}\mathbf{a_{p}}\rangle$ when increasing the inertia of heavy particles of the size of the Kolmogorov-scale.
We found that this increase could not be attributed to the change in the magnitudes of the velocity and/or acceleration increments ($|\delta_{r}\mathbf{v}|$ and/or $|\delta_{r}\mathbf{a_{p}}|$).
A significant contribution to this effect is from the alignment or anti-alignment between the relative velocity and relative acceleration. Using a simple model, \textit{i.e.} a linear velocity field, we show that in the inertial range heavy particles preferentially explore strain-dominated regions of a turbulent flow. 

 The tendency of heavy particles to preferentially explore strain-dominated regions may also explain our recent observations~\citep{Xu:2008p1025} that the time-averaged radial distribution function of particle positions increases when the center of the von K\'arm\'an flow is approached. In that case, the long-time averaged mean flow imposes deterministic strain at the center of the apparatus. With our new findings reported here, one would expect that  the long-time average would lead to a measurable preferential concentration towards the center of the apparatus. 

Our analysis does not rely on the commonly used and hard to measure long time average of the relative particle positions. Instead it only considers dynamical properties and thus opens a novel way to study particle turbulence interaction. For higher Stokes number, the filtering induced by the particle response time~\citep{Bec:2006p656,SalazarCollins:2010} will have an additional effect on the analysis presented above.

We are grateful to R. J. Hill, A. Pumir and E. -W. Saw for many interesting discussions. Support from COST Action MP0806 is kindly acknowledged. This work was funded by the Max Planck Society, and the Marie Curie Fellowship, Programme PEOPLE - Call FP7-PEOPLE-IEF-2008  Proposal N$^{o}$ 237521. 

\bibliographystyle{jfm}
\bibliography{dVdA}

\end{document}